\pdfoutput=1
\documentclass[a4paper,USenglish,cleveref,autoref,thm-restate]{lipics-v2021}
\hideLIPIcs
\nolinenumbers

\usepackage{amsmath,amssymb}
\usepackage{xspace}
\usepackage{booktabs}
\usepackage{tikz}
\usetikzlibrary{calc}

\newcommand{\T}{\mathcal{T}}
\newcommand{\C}{\mathcal{C}}
\newcommand{\K}{\Pi}
\DeclareMathOperator{\env}{env}

\title{Approximate Minimum Triangle Partitions}
\titlerunning{Approximate Minimum Triangle Partitions}

\author{Guilherme D. da Fonseca}{Aix-Marseille Universit\'e and LIS, France \and \url{https://pageperso.lis-lab.fr/guilherme.fonseca/}}{}{https://orcid.org/0000-0002-9807-028X}{}
\authorrunning{G.\,D. da Fonseca}
\Copyright{Guilherme D. da Fonseca}


\EventEditors{}
\EventNoEds{0}
\EventLongTitle{}
\EventShortTitle{}
\EventAcronym{}
\EventYear{}
\EventDate{}
\EventLocation{}
\EventLogo{}
\SeriesVolume{}
\ArticleNo{}

\definecolor{myFill}{HTML}{A0DFFF}
\definecolor{myAlert}{HTML}{c02020}
\definecolor{myDark}{HTML}{5AAFD8}
\tikzset{ray/.style={->,thick,shorten <=1.5pt}, dray/.style={ray,line width=1.4pt,dash pattern=on 3pt off 2pt}}

\begin{document}
\maketitle

\begin{abstract}
Let $\rho(P)$ be the minimum number of triangles in a partition of a simple polygon $P$ with $n$ vertices where triangle corners may be arbitrary points (Steiner points).
It is not known whether $\rho(P)$ can be computed in polynomial time.
We do not assume general position: when vertices of $P$ are collinear, a triangle side may pass through several vertices, and good partitions exploit this.
Every vertex of $P$ is a corner of some triangle of every partition, so $\rho(P)\ge n/3$, and any triangulation, with $n-2$ triangles, is a $3$-approximation that can be computed in $O(n)$ time.
This is the best factor known.

We improve it in two steps.
First, let $\rho_0(P)$ be the minimum number of triangles over partitions whose corners are all vertices of $P$, which can be computed in $O(n^3)$ time (Asano, Asano, and Pinter, 1986).
We prove that $\rho_0(P) \le 2\rho(P)-2$ unless $P$ is a triangle, which is tight.
Thus Steiner points save at most half of the triangles and $\rho_0(P)$ is a $2$-approximation, computed in $O(n^3)$ time.
Second, we break the factor-$2$ barrier of Steiner-free partitions.
We observe that $\rho_0(P)$ is close to $2\rho(P)$ only if optimal partitions have few Steiner points, each with many triangles around it.
We handle this case by a dynamic program over partitions in which every Steiner point in the interior of $P$ is joined to a vertex of $P$ by a straight run of triangle sides, giving a $\frac53$-approximation in $O(n^{12})$ time.
\end{abstract}

\section{Introduction}
\label{sec:intro}

Decomposing a polygon into few simple pieces is a classical topic in computational geometry~\cite{ORST17}, and triangles are the simplest pieces of all.
Every simple polygon with $n$ vertices can be triangulated into $n-2$ triangles in $O(n)$ time~\cite{Chazelle91}.
If the goal is to partition the polygon into as few interior-disjoint triangles as possible, however, triangulations can be wasteful: a triangle side may run through several collinear vertices, and triangle corners may be points that are not vertices of the polygon, called \emph{Steiner points}.
For a polygon $P$ we write:
\begin{itemize}
\item $\rho(P)$ for the minimum number of triangles in a partition of $P$ into triangles;
\item $\rho_0(P)$ for the minimum over \emph{Steiner-free} partitions, whose corners are vertices of $P$.
\end{itemize}

Clearly $\rho(P)\le\rho_0(P)\le n-2$.
We ask how well $\rho_0(P)$, which is computable in $O(n^3)$ time~\cite{AAP86}, approximates $\rho(P)$.
We also introduce an intermediate quantity $\rho_\star(P)$, the minimum size of a partition in a class $\K_\star$ with restricted Steiner points, defined below.

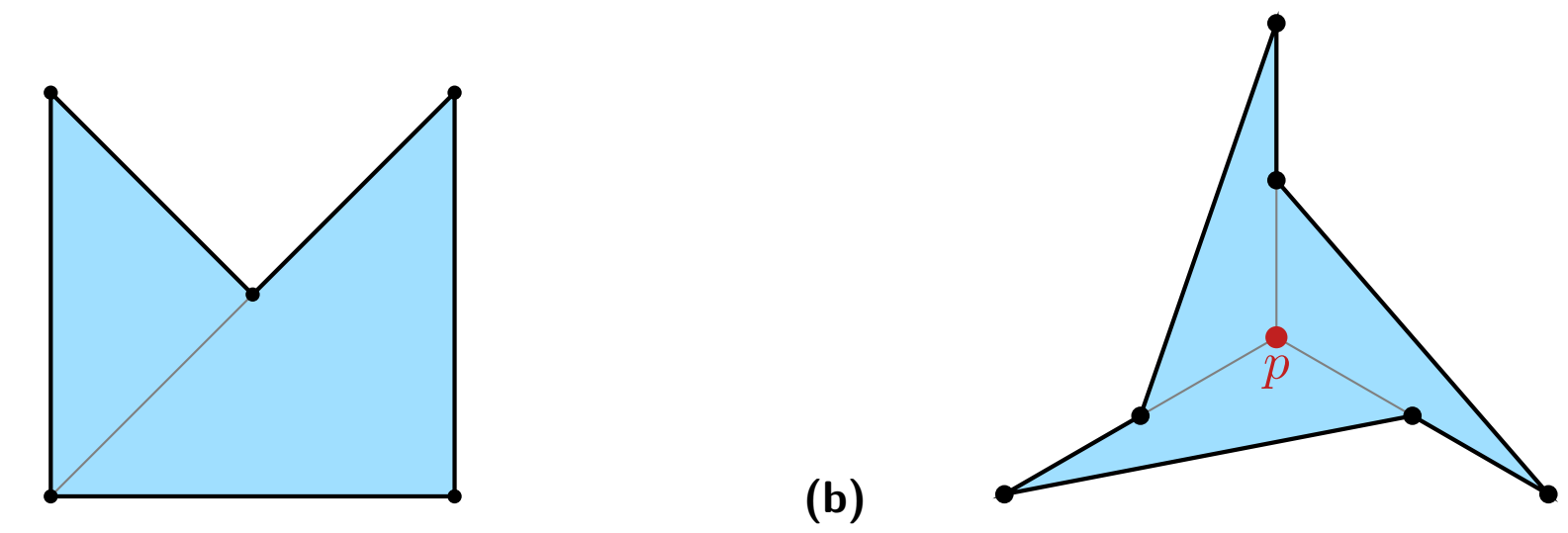
\begin{figure}[t]
\centering
\captionsetup[subfigure]{skip=-5pt}
\begin{subfigure}[b]{0.45\textwidth}
\centering
\begin{tikzpicture}[scale=0.45]
  \fill[myFill] (0,0)--(6,0)--(6,6)--(3,3)--(0,6)--cycle;
  \draw[thick] (0,0)--(6,0)--(6,6)--(3,3)--(0,6)--cycle;
  \draw[gray] (0,0)--(3,3);
  \foreach \q in {(0,0),(6,0),(6,6),(3,3),(0,6)} \fill \q circle (3pt);
\end{tikzpicture}
\caption{}
\label{fig:collinear}
\end{subfigure}
\hfil
\begin{subfigure}[b]{0.45\textwidth}
\centering
\begin{tikzpicture}[scale=0.7]
  \coordinate (p) at (0,0);
  \path (90:3) coordinate (f0) (210:1.5) coordinate (n1) (210:3) coordinate (f1)
        (330:1.5) coordinate (n2) (330:3) coordinate (f2) (90:1.5) coordinate (n0);
  \fill[myFill] (f0)--(n1)--(f1)--(n2)--(f2)--(n0)--cycle;
  \draw[gray] (p)--(f0) (p)--(f1) (p)--(f2);
  \draw[thick] (f0)--(n1)--(f1)--(n2)--(f2)--(n0)--cycle;
  \foreach \q in {f0,n1,f1,n2,f2,n0} \fill (\q) circle (2.5pt);
  \fill[myAlert] (p) circle (3pt) node[below] {$p$};
\end{tikzpicture}
\caption{}
\label{fig:steiner}
\end{subfigure}
\caption{(a)~A pentagon with $\rho_0=2<n-2=3$.
  (b)~A hexagon with no three collinear vertices, so $\rho_0=n-2=4$, partitioned into $3$ triangles using one Steiner point $p$.}
\end{figure}

If no three vertices of $P$ are collinear, then every Steiner-free partition is a triangulation and $\rho_0(P)=n-2$.
With collinear vertices, a triangle side may pass through a reflex vertex, and $\rho_0(P)$ can be much smaller than $n-2$ (\cref{fig:collinear}).
It can be as small as $n/3$, the smallest possible value, since every vertex of $P$ is a corner of some triangle (\cref{lem:corner}).
For example, a triangle with disjoint triangular teeth glued along one of its sides has $\rho_0=n/3$.
Steiner points exploit the same phenomenon: a triangle side with a Steiner endpoint may pass through a vertex of $P$ even if no three vertices are collinear (\cref{fig:steiner}).

Asano, Asano, and Pinter~\cite{AAP86} computed $\rho_0(P)$ in $O(n^3)$ time by dynamic programming, and $\rho_0(P)$ is $n-2$ minus the maximum number of compatible triples of collinear vertices~\cite{CC90}.
Computing $\rho$ is NP-hard for polygons with holes~\cite{Lingas82}, and for simple polygons its complexity is open.
Steiner points at intersections of edge extensions can reduce the number of triangles, and an optimal partition may also need Steiner points on segments joining such intersections~\cite[Section~4]{AAP86}.
Two conjectures followed: for every $k$ some polygon needs $k$ rounds of this construction, and $O(\sqrt n)$ rounds always suffice~\cite[Section~4]{AAP86}.
Kleist and Scherzer~\cite{KS27} refuted the second one in a strong form: there is a simple polygon with rational vertices all of whose minimum partitions have Steiner points with irrational coordinates, which no number of rounds produces.
Abrahamsen, Blikstad, Nusser, and Zhang~\cite{ABNZ26} call the problem ``arguably the most basic'' open partition problem for simple polygons.
The only known approximation factor is $3$~\cite{KS27}: every vertex of $P$ must be a corner of some triangle, so $\rho(P)\ge n/3$, while any triangulation has $n-2$ triangles and can be computed in $O(n)$ time~\cite{Chazelle91}.
For polygons with $h$ holes, the same argument gives the factor $3+6h/n\le5$~\cite{KS27}.
Lower bounds stronger than $n/3$ are known only for special polygons: $\rho(P)=n-2$ if $P$ is convex~\cite{Bankov10}, and more generally if $P$ has at most two reflex vertices and its vertices are in general position~\cite{KS27}.
With three reflex vertices, general position no longer forces $\rho(P)=n-2$~\cite[Figure~3(c)]{KS27}.

\subparagraph{Our results.}\footnote{The previous version of this paper also bounded $\rho_0$ in terms of $\rho$ for polygons with holes, and derived from this bound a $\frac52$-approximation for polygons with a constant number of holes.
The running time of that approximation relied on the algorithm of~\cite{AAP86} for computing $\rho_0$ in polygons with holes.
We were unable to verify the analysis of that algorithm, so we removed this result.}
\begin{enumerate}
\item \emph{Steiner points save at most half.}
  Unless $P$ is a triangle, $\rho_0(P)\le 2\rho(P)-2$ (\cref{thm:main}), and this is tight (\cref{prop:pinwheel}).
  In general position this says $\rho(P)\ge n/2$ for $n\ge4$.
\item \emph{A $5/3$-approximation.}
  We define a class $\K_\star$ of partitions in which every Steiner point lies on two distinct lines, each through two vertices of $P$, and every Steiner point in the interior of $P$ is joined to a vertex of $P$ by a segment that lies on such a line, is covered by triangle sides, and contains no triangle corner in its relative interior.
  A partition with at most $\rho_\star(P)$ triangles, the optimum over $\K_\star$, can be computed in $O(n^{12})$ time (\cref{thm:dp}), and $\rho_\star(P)\le\frac{5\rho(P)-2}{3}$ (\cref{thm:53}).
  The exponent is large, and we make no attempt to reduce it.
\end{enumerate}

\subparagraph{Techniques.} A triangulation uses $n-2$ triangles.
An optimal partition saves triangles in one way only: by \emph{absorbing} vertices of $P$ into the relative interior of triangle sides, at a price for each Steiner point (\cref{lem:angle}).
To obtain a Steiner-free partition, we shrink every maximal straight run of triangle sides to the segment between its first and last vertices of $P$, cut $P$ along these segments, and triangulate the pieces.
This preserves almost every absorption.
The exceptions are absorbed vertices next to a Steiner endpoint of a side, and each Steiner point is responsible for at most as many of them as it has triangles.
Balancing this loss against the trivial bound $\rho_0\le n-2$ gives the factor $2$, with a slack of one triangle per Steiner point and half a triangle per \emph{unproductive} ray, that is, a ray from a Steiner point that is not charged for a lost absorption.

The factor $2$ is approached only when optimal partitions have few Steiner points and almost every ray from a Steiner point absorbs a vertex.
In that regime we keep the Steiner points instead of removing them.
We keep every Steiner point that lies on two lines through pairs of vertices of $P$ and sees a vertex of $P$ along one of them, and we rebuild the triangles around these points.
Each dropped point costs one triangle for every absorption lost next to it, minus its share of the angle count.
Every lost absorption is paid by an unproductive ray, by a Steiner point through which the run of sides passes, or by the share of the dropped point itself, so the cost is at most twice the slack.
The kept points remain joined to vertices of $P$ by straight runs of triangle sides.
This property makes a polynomial dynamic program possible: it extends the dynamic program for $\rho_0$~\cite{AAP86} with candidate Steiner points attached to vertices of $P$.

\subparagraph{Covers.} A related problem allows overlaps.
Let $\kappa(P)$ be the minimum number of triangles contained in $P$ whose union is $P$ (a \emph{cover}), so that $\kappa(P)\le\rho(P)$.
Covers are harder: computing $\kappa$ is $\exists\mathbb{R}$-complete~\cite{Abr21} and APX-hard~\cite{Christ11}.
A $12$-approximation for simple polygons was obtained recently~\cite{KS27}, building on a $6$-approximation for covers by convex polygons~\cite{BKMP23}.
For polygons with holes, an $O(\log n)$-approximation follows from earlier work~\cite{EW03}, as written out in~\cite{KS27}.
Its first step restricts the corners of the triangles to points of $P$ that lie on two lines through pairs of vertices of $P$, which loses a factor at most $25$~\cite[Corollary~33]{KS27}.
The Steiner points of partitions in $\K_\star$ lie on the same points, and there the loss is a factor at most $\frac53$.
If no two edges of $P$ are collinear, then a triangle contains a segment of at most three edges, so $\kappa(P)\ge n/3$ and a triangulation is a $3$-approximation~\cite{Christ11}.
With collinear edges, partitions are of limited use for covers: in the worst case, $\rho(P)/\kappa(P)=\Theta(\alpha(n))$, where $\alpha$ is the inverse Ackermann function.
The upper bound follows from the complexity of a single face in an arrangement of segments~\cite{PSS88,GSS89}: the complement of $P$ lies in one face of the arrangement of the $3\kappa$ sides of a minimum cover, so $n=O(\kappa\,\alpha(\kappa))$.
The lower bound follows from a construction~\cite{Alk25} based on lower envelopes of segments~\cite{WS88}.
We write out the lower bound in \cref{app:covers}.
Hence a triangulation approximates $\kappa$ within a factor $O(\alpha(n))$, but no partition, with or without Steiner points, approximates $\kappa$ within a constant factor, and constant-factor algorithms for covers must output overlapping triangles.

\subparagraph{Organization.} \Cref{sec:prelim} introduces the angle identity on which all our bounds rest.
\Cref{sec:main} proves the factor $2$ and its tightness.
\Cref{sec:beating} defines the class $\K_\star$, builds a near-optimal partition in it, gives the dynamic program, and derives the factor $\frac53$.
\Cref{app:covers} writes out the known lower bound relating partitions and covers.

\section{Preliminaries}
\label{sec:prelim}

This section fixes the notation for partitions and states the angle identity (\cref{lem:angle}), which expresses the size of a partition in terms of its absorbed vertices and its Steiner points.
All our upper and lower bounds are derived from it.
It also states that a region can be triangulated without using its flat corners (\cref{lem:faces}).
These facts are known or follow from classical arguments, and we include short proofs.

Let $P$ be a simple polygon with vertex set $V$, $n=|V|$, and interior angle $\theta_v$ at $v\in V$.
Vertices of $P$ may be collinear, but no three consecutive vertices are, so $\theta_v\ne\pi$.

Let $\T$ be a partition of $P$ into triangles.
A \emph{corner} of $\T$ is a vertex of one of its triangles.
A \emph{Steiner point} of $\T$ is a corner that is not a vertex of $P$.
For a Steiner point $p$, let $c_p$ be the number of triangles of $\T$ with a corner at $p$.
Since triangle angles are less than $\pi$, $c_p\ge2$.
Let $S$ be the set of Steiner points and $c(S)=\sum_{p\in S}c_p$ the number of Steiner corners.

\begin{lemma}[{\cite[Claim~24]{KS27}}]
\label{lem:corner}
Every vertex of $P$ is a corner of $\T$.
Hence $n\le 3|\T|-c(S)$.
\end{lemma}
\begin{proof}
Let $v\in V$.
A triangle containing $v$ has $v$ either as a corner or in the relative interior of a side; in the second case it covers an angle $\pi$ at $v$.
Since $\theta_v<2\pi$ and $\theta_v\ne\pi$, at most one triangle is of the second kind, and some triangle has a corner at $v$.
Counting corners, $3|\T|\ge n+c(S)$.
\end{proof}

For Steiner-free partitions, the resulting bound $\rho_0(P)\ge n/3$ already appears in~\cite[Theorem~3]{CC90}.
A vertex of $P$ is \emph{absorbed} if it lies in the relative interior of a side of a triangle of $\T$.
This side is unique, since a triangle with $v$ in the relative interior of a side covers an angle $\pi$ at $v$, and $\theta_v<2\pi$.
Let $\mu$ be the number of absorbed vertices.
The triangles with a corner at a Steiner point $p$ have angles summing to $s_p\pi$, where $s_p=1$ if $p$ lies in the relative interior of an edge of $P$ (a \emph{boundary point}) or of a side of a triangle (a \emph{T-junction}), and $s_p=2$ otherwise (a \emph{full point}).
If $s_p=1$, the triangles with a corner at $p$ fill a half-plane at $p$, bounded by a line that we call the \emph{carrying line} of $p$.

\begin{lemma}[{Angle identity~\cite[Lemma~9]{KS27}}]
\label{lem:angle}
$|\T|=n-2-\mu+\sum_{p\in S}s_p$.
\end{lemma}
\begin{proof}
Summing all triangle angles by corner gives $|\T|\pi=\sum_{v\in V}\theta_v-\mu\pi+\sum_{p\in S}s_p\pi=(n-2-\mu+\sum_{p\in S}s_p)\pi$.
\end{proof}

The version in~\cite[Lemma~9]{KS27} counts boundary and interior corners and is proved with Euler's formula; there, absorbed vertices and T-junctions are the corners with a flat angle.
For Steiner-free partitions, the corresponding bound in terms of degenerate triangles is~\cite[Lemma~2]{CC90}.
Since a triangulation has no absorbed vertices and no Steiner points, a partition saves exactly $\mu-\sum_{p\in S}s_p$ triangles over a triangulation: one per absorbed vertex, minus the cost $s_p$ of each Steiner point $p$.

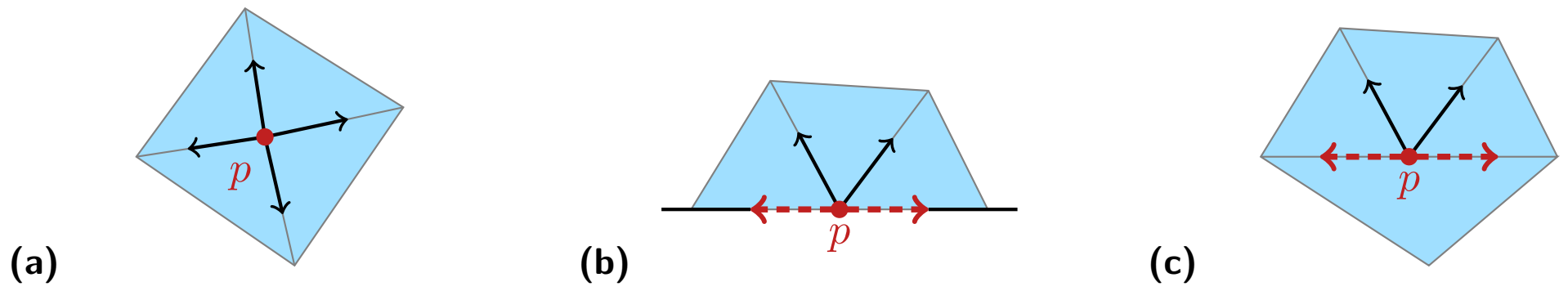
\begin{figure}[t]
\centering
\captionsetup[subfigure]{skip=-5pt}
\begin{subfigure}[b]{0.3\textwidth}
\centering
\begin{tikzpicture}[scale=0.8]
  \coordinate (p) at (0,0);
  \path (1.4,0.3) coordinate (q1) (-0.2,1.3) coordinate (q2) (-1.3,-0.2) coordinate (q3) (0.3,-1.3) coordinate (q4);
  \fill[myFill] (q1)--(q2)--(q3)--(q4)--cycle;
  \draw[gray] (q1)--(q2)--(q3)--(q4)--cycle;
  \foreach \q in {q1,q2,q3,q4} {\draw[gray] (p)--(\q); \draw[ray] (p)--($(p)!0.6!(\q)$);}
  \fill[myAlert] (p) circle (2.5pt); \node[myAlert] at (-0.25,-0.4) {$p$};
\end{tikzpicture}
\caption{}
\end{subfigure}
\hfil
\begin{subfigure}[b]{0.3\textwidth}
\centering
\begin{tikzpicture}[scale=0.8]
  \coordinate (p) at (0,0);
  \path (1.5,0) coordinate (r) (0.9,1.2) coordinate (q1) (-0.7,1.3) coordinate (q2) (-1.5,0) coordinate (l);
  \fill[myFill] (r)--(q1)--(q2)--(l)--cycle;
  \draw[gray] (r)--(q1)--(q2)--(l) (p)--(q1) (p)--(q2);
  \draw[thick] (-1.8,0)--(-0.9,0) (0.9,0)--(1.8,0); \draw[gray] (-0.9,0)--(0.9,0);
  \foreach \q in {q1,q2} \draw[ray] (p)--($(p)!0.6!(\q)$);
  \draw[dray,myAlert] (p)--(0.9,0); \draw[dray,myAlert] (p)--(-0.9,0);
  \fill[myAlert] (p) circle (2.5pt) node[below] {$p$};
\end{tikzpicture}
\caption{}
\end{subfigure}
\hfil
\begin{subfigure}[b]{0.3\textwidth}
\centering
\begin{tikzpicture}[scale=0.8]
  \coordinate (p) at (0,0);
  \path (1.5,0) coordinate (r) (0.9,1.2) coordinate (q1) (-0.7,1.3) coordinate (q2) (-1.5,0) coordinate (l) (0.2,-1.1) coordinate (d);
  \fill[myFill] (r)--(q1)--(q2)--(l)--(d)--cycle;
  \draw[gray] (r)--(q1)--(q2)--(l)--(d)--(r)--(l) (p)--(q1) (p)--(q2);
  \foreach \q in {q1,q2} \draw[ray] (p)--($(p)!0.6!(\q)$);
  \draw[dray,myAlert] (p)--(0.9,0); \draw[dray,myAlert] (p)--(-0.9,0);
  \fill[myAlert] (p) circle (2.5pt) node[below] {$p$};
\end{tikzpicture}
\caption{}
\end{subfigure}
\caption{The rays of (a)~a full point, (b)~a boundary point, and (c)~a T-junction.
  Carrying rays are dashed and red.}
\label{fig:rays}
\end{figure}

\subparagraph{Rays.} A \emph{ray} of a Steiner point $p$ is a direction in which some side with endpoint $p$ leaves $p$.
A full point has $c_p$ rays.
A boundary point or T-junction has $c_p+1$ rays, two of which lie on its carrying line: the two extreme triangles at $p$ have sides on it ending at $p$.
These two rays are \emph{carrying}.
The set $R_p$ of non-carrying rays of $p$ has $|R_p|=c_p-2+s_p$ (\cref{fig:rays}).

\subparagraph{Triangulating regions.} We rebuild partitions by cutting $P$ into regions and triangulating every region without using its flat corners.
Let $F$ be a bounded open region whose boundary is a finite union of segments.
For a point $x$ on the boundary of $F$, the intersection of $F$ with a small disk centered at $x$ consists of finitely many open circular sectors with apex $x$, which we call the \emph{incidences} of $F$ at $x$.
An incidence is \emph{flat} if its angle is $\pi$.
A point may have several incidences, for example at the end of a slit or where the boundary touches itself, and some of them may be flat while others are not.

\begin{lemma}
\label{lem:faces}
Let $F$ be a bounded, open, simply connected region whose boundary is a finite union of segments.
Then the closure of $F$ is the union of interior-disjoint triangles whose interiors lie in $F$ and such that the angle of each triangle at each of its corners lies in a non-flat incidence of $F$.
\end{lemma}
\begin{proof}
For a simple polygon, the lemma follows by deleting the vertices with a flat angle and triangulating.
In general, we follow the proof that every simple polygon has a diagonal~\cite[Theorem~3.1]{BCKO08}, and use induction on the number of non-flat incidences of $F$.
Every vertex of the convex hull of $\bar F$ lies on $\partial F$, and every incidence of $F$ at such a vertex has angle less than $\pi$.
The hull has at least three vertices, so $F$ has at least three non-flat incidences.
Let $w$ be a hull vertex and consider an incidence of $F$ at $w$.
Its two bounding rays lie in $\partial F$.
Follow each of them from $w$ as long as the point lies on $\partial F$ and $F$ contains a half-disk at it on the side of the incidence, and let $a$ and $b$ be the points where this stops.
Let $T$ be the triangle $awb$.
Then $F$ contains the points of $T$ near the half-open segments $[w,a)$ and $[w,b)$, and the incidences of $F$ at $a$ and at $b$ on the side of $T$, next to $[w,a)$ and $[w,b)$, respectively, are not flat.
Let $H$ be the set of points of $\partial F\cap T$ that do not lie on $[w,a)\cup[w,b)$; it is compact and contains $a$ and $b$.

If some point of $H$ lies off the line $ab$, let $\ell$ be the line parallel to $ab$ through a point of $H$ farthest from $ab$, and let $z$ be an endpoint of a connected component of $H\cap\ell$.
The part of the interior of $T$ beyond $\ell$ on the side of $w$ contains no point of $\partial F$ and meets $F$, so it lies in $F$.
Hence the open segment $wz$ lies in $F$, and so does the open half-disk at $z$ on the side of $w$.
The point $z$ is interior to $T$, since it lies off $ab$ and not on $[w,a)\cup[w,b)$.
The incidence of $F$ at $z$ that contains this half-disk is not flat: otherwise $\ell$ would lie in $\partial F$ on both sides of $z$, and these points, interior to $T$, would extend $H\cap\ell$ beyond $z$.
We cut $F$ along $wz$.

Otherwise $H$ lies on the line $ab$, and the interior of $T$, which contains no point of $\partial F$ and meets $F$, lies in $F$.
If the open segment $ab$ lies in $\partial F$, then $F$ is the interior of $T$, and we are done.
If it misses $\partial F$, it lies in $F$, and we cut $F$ along $ab$.
Otherwise some point $z$ of the open segment $ab$ is an endpoint of a connected component of $\partial F\cap ab$.
The points of $ab$ next to $z$ on one side lie in $F$, so the incidence of $F$ at $z$ that contains the interior of $T$ near $z$ has angle more than $\pi$, and we cut $F$ along the open segment $wz$, which lies in $F$.

In each case the cut is an open segment in $F$ whose endpoints lie on $\partial F$ in non-flat incidences that contain the cut near them.
Since $F$ is simply connected, the cut splits $F$ into two simply connected regions $F_1$ and $F_2$ bounded by finitely many segments.
Their incidences are those of $F$, except that the incidences at the endpoints of the cut are split in two, and the points of the cut get flat incidences.
Hence the flat incidences of $F$ remain flat, every non-flat incidence of $F_1$ or $F_2$ is contained in a non-flat incidence of $F$, and $F_1$ and $F_2$ together have at most two more non-flat incidences than $F$.
Since each of them has at least three, each has fewer than $F$, and we apply induction to $F_1$ and $F_2$.
\end{proof}

\subparagraph{Candidate points.} A \emph{vertex-pair line} is a line through two vertices of $P$.
Let $\C$ be the set of points of $P$ that lie on two distinct vertex-pair lines; $|\C|=O(n^4)$.
The points of $\C$ are the possible Steiner points of the partitions computed by our algorithm.

\section{Steiner points save at most half}
\label{sec:main}

In this section we prove that $\rho_0(P)\le2\rho(P)-2$ unless $P$ is a triangle (\cref{thm:main}).
Intuitively, compared with a triangulation, a partition saves one triangle for every vertex of the polygon that lies in the middle of a triangle side, and pays one or two triangles for every Steiner point of an optimal partition $\T$.
We start from $\T$ and follow the lines formed by its triangle sides.
On each line, we keep only the stretch from the first vertex of the polygon to the last.
These stretches are non-crossing diagonals, so we can cut along them and triangulate the pieces without Steiner points, skipping the vertices that lie flat along the cuts.
Every saving of $\T$ survives, except at a vertex that is the outermost one on its line, next to a Steiner point.
Each such lost saving can be blamed on a side leaving a Steiner point.
So the Steiner-free partition exceeds the optimum by at most the number of triangle corners at Steiner points.
On the other hand, every vertex of the polygon is a corner of some triangle, so a triangulation has at most three times as many triangles as the optimum, minus those same Steiner corners.
The first bound is good when Steiner points have few corners and the second when they have many; the better of the two is at most twice the optimum.
Pinwheels, in which one central Steiner point has a saving on every side leaving it, show that no Steiner-free partition does better than twice the optimum.

\subparagraph{Chords.} Fix an optimal partition $\T$ with Steiner points $S$.
On every line, the union of the sides of $\T$ contained in it consists of segments.
At a full point where sides pass straight through along two or more lines, we split these segments at the point on all of these lines but at most one, chosen once and for all (the \emph{merging rule}); the resulting segments are the \emph{components}.
For a component $K$ containing at least two vertices of $P$, its \emph{chord} $\hat K$ is the segment between the first and the last vertex of $P$ on $K$.
Every chord joins two vertices of $P$ and lies in $P$, since $\hat K\subseteq K\subseteq P$.

\begin{lemma}
\label{lem:chords}
A segment between two consecutive vertices of $P$ on a chord is an edge or a diagonal of $P$, and no two such segments cross.
\end{lemma}
\begin{proof}
Let $uw$ be such a segment.
It lies in $P$ and has no vertex of $P$ in its relative interior.
If it touches the relative interior of an edge, it cannot cross the edge, so it runs along it near the contact point.
Since its endpoints are vertices, it equals the edge.
Otherwise it is a diagonal.

Two sides of a partition do not cross.
Suppose that two such segments on different lines cross at a point $x$.
The point $x$ is not a vertex of $P$, since the segments contain no vertex in their relative interiors.
Hence some sides pass straight through $x$ along one of the lines.
If $x$ were in the relative interior of a side $f$ on the other line, the continuation through $x$ would enter the interior of the triangle bounded by $f$, which contains no side.
Hence sides pass straight through $x$ along both lines, so $x$ is a full Steiner point, and by the merging rule one of the two components ends at $x$, a contradiction.
\end{proof}

By \cref{lem:chords}, cutting $P$ along all chords splits it by non-crossing diagonals into simple polygons, which we call \emph{faces}.
Their corners are vertices of $P$.
A corner of a face is \emph{flat} if the face has angle $\pi$ there, and $\Phi$ denotes the total number of flat corners of the faces.
Triangulating each face without using its flat corners gives a Steiner-free partition, whose size is given by an angle count.

\begin{lemma}
\label{lem:count}
The resulting partition has at most $n-2-\Phi$ triangles.
\end{lemma}
\begin{proof}
The angles of all faces at all their vertices add up to $\sum_v\theta_v=(n-2)\pi$.
A face $F$ with $n_F$ vertices, $\Phi_F$ of them flat, has angle sum $(n_F-2)\pi$, and the angles of its triangles lie in its other corners, so it yields at most $n_F-2-\Phi_F$ triangles.
\end{proof}

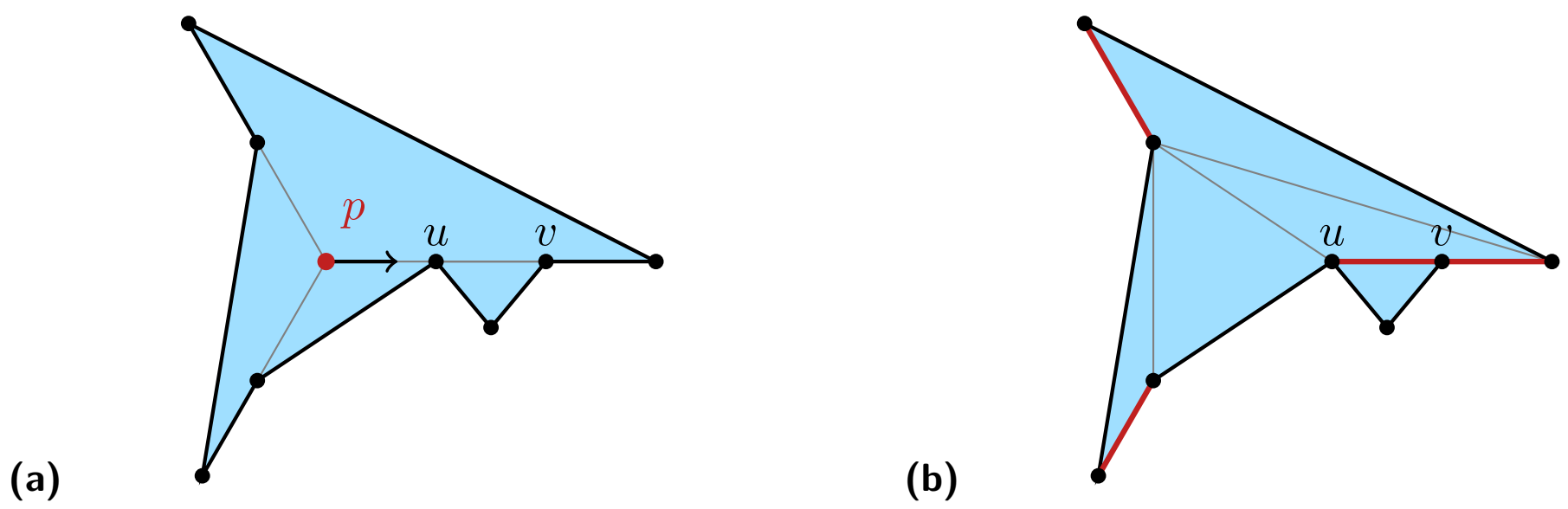
\begin{figure}[t]
\centering
\captionsetup[subfigure]{skip=-5pt}
\newcommand{\chordpoly}{
  \path (0,0) coordinate (p) (6,0) coordinate (a0) (120:5) coordinate (b1) (120:2.5) coordinate (a1) (240:4.5) coordinate (b2) (240:2.5) coordinate (a2) (2,0) coordinate (u) (3,-1.2) coordinate (y) (4,0) coordinate (v); \fill[myFill] (a0)--(b1)--(a1)--(b2)--(a2)--(u)--(y)--(v)--cycle; }
\begin{subfigure}[b]{0.45\textwidth}
\centering
\begin{tikzpicture}[scale=0.42]
  \chordpoly
  \draw[gray] (p)--(a0) (p)--(b1) (p)--(b2) (u)--(v);
  \draw[thick] (a0)--(b1)--(a1)--(b2)--(a2)--(u)--(y)--(v)--cycle;
  \foreach \q in {a0,b1,a1,b2,a2,u,y,v} \fill (\q) circle (4pt);
  \draw[ray] (p)--(1.3,0);
  \fill[myAlert] (p) circle (4.5pt); \node[myAlert] at (60:1) {$p$};
  \node[above] at (u) {$u$}; \node[above] at (v) {$v$};
\end{tikzpicture}
\caption{}
\end{subfigure}
\hfil
\begin{subfigure}[b]{0.45\textwidth}
\centering
\begin{tikzpicture}[scale=0.42]
  \chordpoly
  \draw[gray] (a0)--(a1)--(u) (a1)--(a2);
  \draw[thick] (a0)--(b1)--(a1)--(b2)--(a2)--(u)--(y)--(v)--cycle;
  \draw[very thick,myAlert] (u)--(a0) (a1)--(b1) (a2)--(b2);
  \foreach \q in {a0,b1,a1,b2,a2,u,y,v} \fill (\q) circle (4pt);
  \node[above] at (u) {$u$}; \node[above] at (v) {$v$};
\end{tikzpicture}
\caption{}
\end{subfigure}
\caption{(a)~A partition with one Steiner point $p$.
  On the component along the arrow, $u$ is lost and $v$ is safe.
  (b)~The chords (red) and a Steiner-free partition that does not use the flat corner $v$.}
\label{fig:chords}
\end{figure}

\subparagraph{Safe and lost absorptions.} It remains to show that $\Phi$ is close to $\mu$.
Let $v$ be a vertex absorbed in a side of a triangle $\Delta$, and let $K$ be the component containing this side.
We say $v$ is \emph{safe} if it is not the first or last vertex of $P$ on $K$, and \emph{lost} otherwise (\cref{fig:chords}).
Let $L$ be the number of lost vertices.

\begin{lemma}
\label{lem:kept}
For each safe vertex $v$, the face containing a small half-disk around $v$ on the side of $\Delta$ has a flat incidence at $v$.
Hence $\Phi\ge \mu-L$, where $\mu$ is the number of absorbed vertices.
\end{lemma}
\begin{proof}
The vertex $v$ lies in the relative interior of $\hat K$.
The half-disk lies in the interior of $\Delta$, which meets no chord, so a single face contains it, and the incidence of this face at $v$ that contains the half-disk is the half-disk itself, which is flat.
Distinct safe vertices give distinct flat incidences, which in a simple face are flat corners.
\end{proof}

Lost vertices are charged to rays of Steiner points.
An \emph{end ray} of a Steiner point $p$ is a ray of $p$ whose component ends at $p$.

\begin{lemma}
\label{lem:lost}
Every end ray is non-carrying.
Every lost vertex $v$ is the first vertex of $P$ seen along some end ray, and $v$ is determined by that end ray.
Hence $L\le\sum_{p\in S}|R_p|$, where $R_p$ is the set of non-carrying rays of $p$.
\end{lemma}
\begin{proof}
At a boundary point or T-junction, sides continue through $p$ along the carrying line, so carrying rays are not end rays.
Let $v$ be lost, absorbed in a side on the component $K$.
The component continues beyond $v$ on both sides, and on one side it contains no further vertex of $P$.
The endpoint $e$ of $K$ on that side is a corner and not a vertex, so it is a Steiner point, $K$ ends at $e$, and $v$ is the vertex of $P$ on $K$ nearest to $e$.
\end{proof}

Assign each lost vertex $v$ to an end ray as in \cref{lem:lost}, namely the end ray at the end of the component of $v$ from which $v$ is the first vertex of $P$; the assignment is injective.
A ray in $R_p$ is \emph{productive} if a lost vertex is assigned to it and \emph{unproductive} otherwise.
Let $\lambda$ be the number of unproductive rays, so that $L=\sum_p|R_p|-\lambda$.

\subparagraph{The factor 2.} Combining the lemmas bounds $\rho_0$ in terms of an optimal partition.
This bound is exact for pinwheels (\cref{sec:tight}).

\begin{lemma}
\label{lem:refined}
For every optimal partition of a simple polygon $P$ with Steiner points $S$,
\[
\rho_0(P)\;\le\;\rho(P)+\sum_{p\in S}(c_p-2)-\lambda,
\]
where $c_p$ is the number of triangles with a corner at $p$.
\end{lemma}
\begin{proof}
By \cref{lem:count,lem:kept,lem:lost,lem:angle}, $\rho_0(P)\le n-2-\Phi\le n-2-\mu+L=n-2-\mu+\sum_p(s_p+c_p-2)-\lambda=\rho(P)+\sum_p(c_p-2)-\lambda$.
\end{proof}

\begin{theorem}
\label{thm:main}
For every simple polygon $P$ and every optimal partition of $P$ with Steiner points $S$ and $\lambda$ unproductive rays,
\[
\rho_0(P)\;\le\;2\rho(P)-1-|S|-\lambda/2.
\]
Moreover, $\rho_0(P)\le2\rho(P)-2$ if $P$ is not a triangle.
Hence the dynamic program for $\rho_0$~\cite{AAP86} is a $2$-approximation for $\rho$ that runs in $O(n^3)$ time.
\end{theorem}
\begin{proof}
\Cref{lem:refined} gives $\rho_0\le\rho+c(S)-2|S|-\lambda$, and a triangulation together with \cref{lem:corner} gives $\rho_0\le n-2\le 3\rho-c(S)-2$.
The smaller of the two bounds is at most their average.
If $P$ is not a triangle, then $\rho\ge2$.
If $S=\emptyset$, the optimal partition is Steiner-free and $\rho_0=\rho\le2\rho-2$.
Otherwise $|S|\ge1$ and the averaged bound $2\rho-1-|S|-\lambda/2$ gives $\rho_0\le2\rho-2$.
\end{proof}

Thus $\rho_0$ can be close to $2\rho$ only if the optimal partitions have few Steiner points and almost all their rays are productive.
The two averaged bounds must then be close as well, so $c(S)\approx\rho$, and the few Steiner points have many corners each.
\Cref{sec:beating} handles this case.
The rest of this section shows that the factor $2$ is tight for Steiner-free partitions.

\subsection{Tightness for Steiner-free partitions: pinwheels}
\label{sec:tight}

We now show that \cref{thm:main} is tight: no algorithm that outputs a Steiner-free partition approximates $\rho$ within a factor better than $2$.
The extremal polygons are pinwheels, in which a single Steiner point absorbs one vertex on each of its rays.

\begin{figure}[t]
\centering
\begin{tikzpicture}[scale=0.55]
  \coordinate (p) at (0,0);
  \fill[myFill] (4,-2)--(6,-3)--(6,2)--(3,1)--(-1,0)--(-4,0)--(0,-2.5)--(0,-3.5)--cycle;
  \draw[gray] (p)--(6,-3) (p)--(6,2) (p)--(3,1) (p)--(-4,0) (p)--(0,-3.5);
  \draw[thick] (4,-2)--(6,-3)--(6,2)--(3,1)--(-1,0)--(-4,0)--(0,-2.5)--(0,-3.5)--cycle;
  \foreach \q in {(4,-2),(6,-3),(6,2),(3,1),(-1,0),(-4,0),(0,-2.5),(0,-3.5)} \fill \q circle (2.5pt);
  \fill[myAlert] (p) circle (3pt) node[below left] {$p$};
\end{tikzpicture}
\caption{A pinwheel with four triangles around the Steiner point $p$.
  The polygon has eight vertices, no three of them collinear, so $\rho_0=6$, whereas $\rho=4$.
  On each ray from $p$, the longer of the two triangle sides absorbs the nearer vertex.}
\label{fig:pinwheel}
\end{figure}
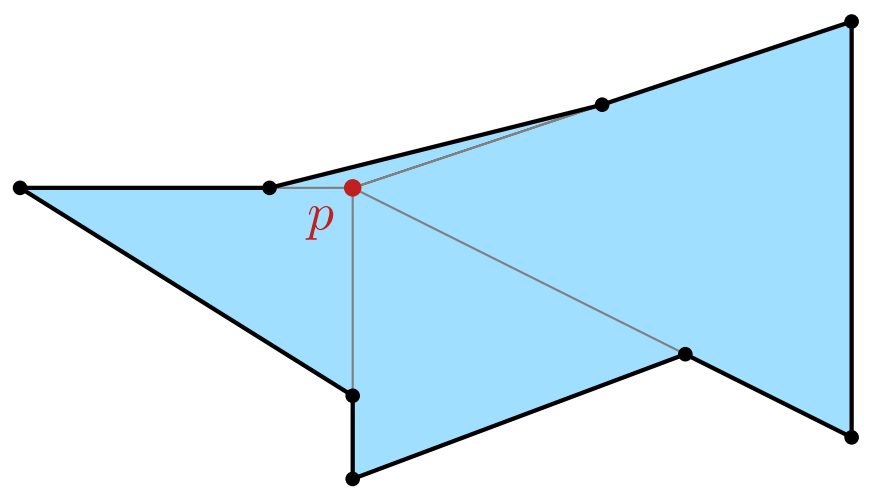

A \emph{pinwheel} with $m\ge3$ triangles is the union of triangles $(p,\;\ell_ig_i,\;\ell'_{i+1}g_{i+1})$ for $i\in\mathbb{Z}_m$.
Here $p$ is the origin, $g_0,\dots,g_{m-1}$ are unit vectors in counterclockwise order with consecutive angles less than $\pi$, and $\ell_i,\ell'_i>0$ with $\ell_i\ne\ell'_i$ for every $i$ (\cref{fig:pinwheel}).
The two triangles on the ray of $g_i$ end at different distances, so the ray carries two vertices, and the nearer one is absorbed by the longer side.
Hence a pinwheel has $n=2m$ vertices and $\rho\le m$.

\begin{proposition}
\label{prop:pinwheel}
For every $m\ge3$ there is a pinwheel with $\rho=m$ and $\rho_0=2m-2=2\rho-2$.
\end{proposition}
\begin{proof}
Choose directions and lengths generically, so that no three vertices are collinear.
Then no vertex can be absorbed in a Steiner-free partition, and \cref{lem:angle} with $\mu=0$ and $S=\emptyset$ shows that every Steiner-free partition has exactly $n-2=2m-2$ triangles.
\Cref{thm:main} gives $\rho\ge(\rho_0+2)/2=m$, so $\rho=m$.
\end{proof}

The pinwheels of \cref{prop:pinwheel} attain \cref{lem:refined} and the bound $\rho_0\le2\rho-2$ of \cref{thm:main} with equality, and they show that the factor $2$ cannot be improved without Steiner points.
The center of a pinwheel lies in $\C$, since each of its rays carries two vertices.
The segment from the center to the nearer of these vertices is a link in the sense of \cref{sec:beating}.
Hence the pinwheel partition belongs to the class $\K_\star$ defined there, and $\rho_\star=\rho$ for the pinwheels of \cref{prop:pinwheel}.

\section{Beating the factor 2}
\label{sec:beating}

In this section we beat the factor $2$ of \cref{thm:main} by keeping Steiner points instead of removing them.
Intuitively, the chord construction loses a factor close to $2$ only when an optimal partition $\T$ has few Steiner points and almost every side leaving them is blamed for a lost saving.
So instead of removing all Steiner points, we keep those that lie where two lines through pairs of vertices cross and that are joined by a side to a vertex along one of these lines.
We rebuild the partition as before, now also cutting at the kept points, and each kept point stays joined to its vertex.
A lost saving is now blamed on a side leaving a dropped Steiner point, and each such side is paid for in one of three ways: by a side leaving a Steiner point that carries no saving, by a Steiner point that the line of the side passes through, or by the triangles saved when its Steiner point was dropped.
So the rebuilt partition exceeds the optimum by at most twice the slack that the chord construction leaves below twice the optimum.
To find a partition no larger than the best one of this kind, we extend the dynamic program for Steiner-free partitions, replacing pairs of vertices by pairs of points that are either vertices or candidate points joined to a vertex.
The segments that join kept points to vertices connect every Steiner point to the boundary, so the triangles form a tree that the program can follow.
The first construction gives twice the optimum minus the slack, the second gives the optimum plus at most twice the slack, and the better of the two is at most five thirds of the optimum.

\subparagraph{The class.} Let $\T'$ be a triangle partition of $P$.
A \emph{link} of a Steiner point $r$ of $\T'$ is a segment $ru$, where $u$ is a vertex of $P$, such that the line through $r$ and $u$ is a vertex-pair line, $ru$ is contained in the union of the sides of $\T'$, and no corner of $\T'$ lies in the relative interior of $ru$.
Let $\K_\star$ be the class of triangle partitions of $P$ such that (i) every Steiner point lies in $\C$, and (ii) every Steiner point in the interior of $P$ has a link.
Let $\rho_\star(P)$ be the minimum size of a partition in $\K_\star$.
Steiner points may lie in the relative interior of sides, and triangles may have two or three Steiner corners.
Steiner-free partitions belong to $\K_\star$, so $\rho\le\rho_\star\le\rho_0$.
The inequality $\rho\le\rho_\star$ can be strict: points of $\C$ have rational coordinates if the vertices of $P$ do, and there is a simple polygon with rational vertices all of whose minimum partitions have Steiner points with irrational coordinates~\cite[Theorem~2]{KS27}.

\Cref{sec:keep} rebuilds $\T$ around kept Steiner points and charges the cost to rays of dropped points.
\Cref{sec:linked} chooses the kept points, shows that the result is in $\K_\star$, and pays for charged rays.
\Cref{sec:dp} gives the dynamic program, and \cref{sec:factor} combines the bounds.

\subsection{Keeping Steiner points}
\label{sec:keep}

We now generalize the chord construction of \cref{sec:main} to partitions that keep a prescribed set of Steiner points.
For an empty set of kept points, the construction below is exactly the chord construction.
In general, the rebuilt partition pays one triangle for each absorbed vertex or kept T-junction that does not stay in the relative interior of a side, and it saves $s_q$ triangles for each dropped Steiner point $q$.

Let $M\subseteq S$ be a set of Steiner points of the optimal partition $\T$ to keep; the points of $S\setminus M$ are \emph{dropped}.
A ray of a Steiner point $p$ is \emph{vertex-bearing} if its component beyond $p$ contains a vertex of $P$.
We assume that every kept point lies on $\partial P$ or has a vertex-bearing ray.

\subparagraph{Construction.} On every component, the \emph{anchors} are its vertices of $P$ and its kept Steiner points.
A \emph{cut} is a segment between two consecutive anchors of the same component.
The \emph{faces} are the connected components of the interior of $P$ minus the union of the cuts.
Faces are simply connected (\cref{lem:keep} below), and we triangulate every face without using its flat incidences, by \cref{lem:faces}; we call the result $\Psi(M)$.

\subparagraph{Charged rays.} An absorbed vertex of $P$ lies in the relative interior of a side, and so does a T-junction; we call the component of this side the component of the point.
Let $x$ be an absorbed vertex or a kept T-junction, and $K$ its component.
We say that $x$ is \emph{exposed} on a side if $K$ has no anchor on that side of $x$.
Every exposed point $x$ \emph{charges} the ray toward $x$ from the end of $K$ on such a side; if $x$ is exposed on both sides, it charges one of the two rays.
Let $\chi$ be the number of charged rays.

\begin{lemma}[Keep-set lemma]
\label{lem:keep}
The cuts do not cross, every face is simply connected, and every Steiner point of $\Psi(M)$ lies in $M$.
Every charged ray is an end ray of a dropped point, the first anchor on the ray is the point that charges it, and no ray is charged twice.
Moreover,
\[
|\Psi(M)|\;\le\;\rho+\chi-\sum_{q\in S\setminus M}s_q.
\]
\end{lemma}
\begin{proof}
Cuts lie in the union of the sides, so, as in the proof of \cref{lem:chords}, two cuts on different lines could only cross at a point where sides continue straight through along both lines.
Such a point is not a vertex of $P$, since vertices are anchors and no anchor lies in the relative interior of a cut.
So it is a full Steiner point, since at boundary points and T-junctions sides continue straight only along the carrying line; by the merging rule one of the two components ends there, so no cut of it contains the point in its relative interior.
So cuts do not cross.
Every kept point in the interior of $P$ is joined to a vertex by consecutive cuts along a vertex-bearing ray, so the union of $\partial P$ and the cuts is connected, and every face is simply connected, though possibly weakly simple, with slits at kept points.
A dropped point is never an endpoint of a cut, and a cut ends in the relative interior of an edge of $P$ only at a kept boundary point.
Hence every point of the boundary of a face that is not in $V\cup M$ lies in the relative interior of a cut or an edge, and all its incidences are flat.
By \cref{lem:faces}, every corner of $\Psi(M)$ lies in $V\cup M$.

Let $x$ be exposed on a side of its component $K$, and let $e$ be the end of $K$ on that side.
Then $e$ is a corner of $\T$ that is neither a vertex nor a kept point, so $e$ is dropped.
Since $K$ ends at $e$, the ray of $e$ that points to $x$ is an end ray, and its first anchor is $x$.
A charged ray determines its first anchor, so no ray is charged twice.

Let $\mu'$ be the number of absorbed vertices of $\Psi(M)$ and $s'_p$ the angle sum of its triangles at a Steiner point $p$, divided by $\pi$.
Let $S'\subseteq M$ be the set of Steiner points of $\Psi(M)$.
By \cref{lem:angle}, applied to $\T$ and to $\Psi(M)$,
\[
|\Psi(M)|-\rho=(\mu-\mu')+\sum_{p\in S'}(s'_p-s_p)-\sum_{p\in S\setminus S'}s_p,
\]
and the last sum is at least $\sum_{q\in S\setminus M}s_q$.
We show that $\mu-\mu'$ is at most the number of exposed absorbed vertices and $s'_p-s_p$ is at most $1$ for exposed T-junctions in $S'$ and at most $0$ for other points of $S'$.
Since every exposed point charges one ray, this proves the bound.

\emph{Absorbed vertices.} Let $u$ be absorbed in a side of a triangle $\Delta_u$ of $\T$, and assume that $u$ is not exposed.
Then the cuts cover a neighborhood of $u$ on its component, and the half-disk at $u$ inside $\Delta_u$ meets no cut.
So this half-disk lies in a flat incidence of a face at $u$.
By \cref{lem:faces}, no triangle of $\Psi(M)$ has its angle at $u$ in this incidence, so $u$ lies in the relative interior of a side of $\Psi(M)$, that is, $u$ is absorbed in $\Psi(M)$.

\emph{Kept points.} Let $p\in M$ be a Steiner point of $\Psi(M)$.
If $p$ is a full point of $\T$, then $s'_p\le2=s_p$, and if $p$ is a boundary point, then $s'_p=1=s_p$.
Let $p$ be a T-junction of $\T$, carried by a side of a triangle $\Delta_p$, so that $s_p=1$ and $s'_p\le2$.
If $p$ is not exposed, then as above the half-disk at $p$ inside $\Delta_p$ lies in a flat incidence of a face, $p$ lies in the relative interior of a side of $\Psi(M)$, and $s'_p=1$.
\end{proof}

Charged rays are end rays of dropped points, so they lie in the sets $R_q$ by \cref{lem:lost}.
Hence $\chi\le\sum_{q\in S\setminus M}|R_q|$, and \cref{lem:keep} gives $|\Psi(M)|\le\rho+\sum_{q\in S\setminus M}(c_q-2)$; for $M=\emptyset$ this is a weak form of \cref{lem:refined}.
The next subsection bounds $\chi$ more carefully.

\subsection{Linked points}
\label{sec:linked}

We now choose the kept points: the points of $\C$ that see a vertex of $P$ along a vertex-pair line.
Their links survive the construction of \cref{sec:keep}, so the rebuilt partition lies in $\K_\star$, and we pay for its charged rays with $\lambda+|S|$, at most twice the slack of \cref{thm:main}.

The \emph{first corner} of a ray of a Steiner point $p$ is the corner of $\T$ on the ray nearest to $p$.
It exists since a side leaves $p$ along the ray, and the segment from $p$ to the first corner lies in this side.
A Steiner point $r$ in the interior of $P$ is \emph{linked} if one of its rays lies on a vertex-pair line and has a vertex $u$ of $P$ as its first corner; then $ru$ is a link of $r$ in $\T$.
From now on, we fix
\[
M=\{r\in S\cap\C : r\in\partial P\text{ or $r$ is linked}\}.
\]
A linked point has a vertex-bearing ray, so \cref{lem:keep} applies to $M$.

\begin{lemma}
\label{lem:links}
The partition $\Psi(M)$ belongs to $\K_\star$.
\end{lemma}
\begin{proof}
By \cref{lem:keep}, every Steiner point of $\Psi(M)$ lies in $M\subseteq\C$, so (i) holds.
Let $r$ be a Steiner point of $\Psi(M)$ in the interior of $P$.
Then $r$ is linked: some ray of $r$ lies on a vertex-pair line and has a vertex $u$ of $P$ as its first corner.
The points $r$ and $u$ are anchors of the component of this ray, and no corner of $\T$ lies between them, so they are consecutive anchors and $ru$ is a cut.
Every cut lies on the boundary of a face, which is covered by the sides of the triangles of that face, so $ru$ lies in the union of the sides of $\Psi(M)$.
The corners of $\Psi(M)$ lie in $V\cup M$ by \cref{lem:keep}, and these are corners of $\T$ by \cref{lem:corner}, so no corner of $\Psi(M)$ lies in the relative interior of $ru$.
Hence $ru$ is a link of $r$ in $\Psi(M)$, and (ii) holds.
\end{proof}

We now bound the number of charged rays.
A component \emph{passes} a Steiner point $x$ if $x$ lies in the relative interior of the component.
At most one component passes $x$: if $x$ is a boundary point or a T-junction, sides pass straight through $x$ only along its carrying line, since on the other side of this line there is the exterior of $P$ or the interior of the triangle whose side carries $x$; if $x$ is a full point, this is the merging rule.

\begin{figure}[t]
\centering
\begin{tikzpicture}[scale=0.8]
  \path (0,0) coordinate (q) (2,0) coordinate (u) (4,0) coordinate (e) (2,1.5) coordinate (t)
        (0.7,-1.2) coordinate (s1) (3.3,-1.2) coordinate (s2);
  \fill[myFill] (q)--(e)--(t)--cycle (q)--(u)--(s1)--cycle (u)--(e)--(s2)--cycle;
  \draw[gray] (q)--(e)--(t)--cycle (q)--(s1) (e)--(s2);
  \draw[gray] (q)--(-1.2,0.9) (q)--(-1.3,-0.6) (e)--(5.2,0.9) (e)--(5.3,-0.6);
  \draw[thick] (s1)--(u)--(s2);
  \draw[ray] (q)--(1.2,0); \draw[dray] (e)--(2.8,0);
  \fill (u) circle (2pt) node[above] {$v$};
  \fill[myAlert] (q) circle (2.5pt) node[above=1pt] {$q$};
  \fill[myAlert] (e) circle (2.5pt) node[above=1pt] {$e$};
  \node[below] at (1,0) {$K$};
\end{tikzpicture}
\caption{A charged ray of type~2 in the proof of \cref{lem:charged}: the component $K$ has the single vertex $v$.
  The productive ray of $q$ has the unproductive partner at $e$ (dashed).}
\label{fig:deep}
\end{figure}
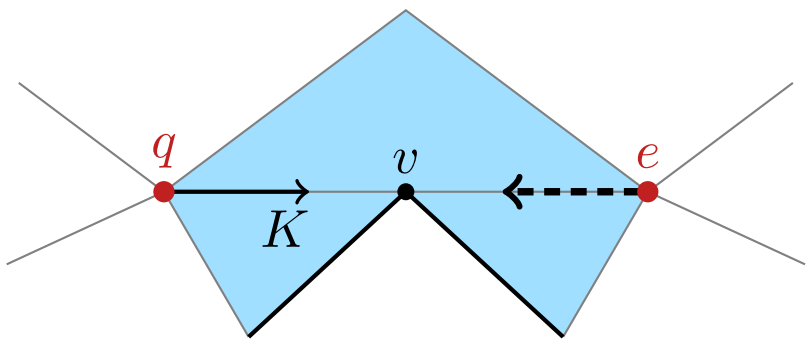

Intuitively, every charged ray is an end ray of a dropped point $q$, and if it is productive, its component contains the vertex assigned to it.
If the line of the ray contains no other vertex, the component continues beyond this vertex to a Steiner point whose ray back is unproductive, as in \cref{fig:deep}.
If the line is a vertex-pair line, then either a Steiner point lies before the vertex, or $q$ sees the vertex directly; in the second case $q$ is not in $\C$, since otherwise it would be linked and kept.
Such a ray is paid by the term $s_q$.

\begin{lemma}
\label{lem:charged}
For the set $M$ defined above, $\chi\le\lambda+|S|+\sum_{q\in S\setminus M}s_q$.
\end{lemma}
\begin{proof}
Let $\eta$ be a charged ray.
By \cref{lem:keep}, it is an end ray of a dropped point $q$, so it lies in $R_q$ (\cref{lem:lost}); let $K$ be its component, which ends at $q$.
We distinguish four types.

\emph{Type 1: $\eta$ is unproductive.}

\emph{Type 2: $\eta$ is productive and its line is not a vertex-pair line.}
Let $v$ be the lost vertex assigned to $\eta$.
Then $v$ is the only vertex of $P$ on $K$, and $v$ is absorbed in a side on $K$.
Hence $K$ continues beyond $v$ to its other end $e$, which is a corner and not a vertex, so $e\in S$.
The ray of $e$ that points to $v$ is an end ray and therefore lies in $R_e$.
The only vertex it could have been assigned is $v$, which is assigned to $\eta$, so the ray of $e$ is unproductive.
We call it the \emph{partner} of $\eta$ (\cref{fig:deep}).
Since $K$ has two ends, distinct rays of type~2 have distinct partners.
We say that $\eta$ has type~2a if its partner is not charged, and type~2b otherwise.
In type~2b, both $\eta$ and its partner are charged, each by its first anchor (\cref{lem:keep}).
The anchors on $K$ are $v$ and kept points, and $v$ charges only one ray, so $K$ passes a kept point.

\emph{Type 3: $\eta$ is productive, its line is a vertex-pair line, and its first corner $x$ is a Steiner point.}
The vertex assigned to $\eta$ is the first vertex of $P$ on $K$ beyond $q$, and it is a corner, so $x$ lies strictly between $q$ and this vertex, and $K$ passes $x$.

\emph{Type 4: $\eta$ is productive, its line is a vertex-pair line, and its first corner is a vertex of $P$.}
Then $q\notin\C$: otherwise $q$ would lie on $\partial P$ or be linked, and $q$ would be kept.
So the line of $\eta$ is the only vertex-pair line through $q$.
A full point has at most two rays on a line, and a boundary point or T-junction has at most one non-carrying ray on a line, since these rays lie in an open half-plane.
Hence $q$ has at most $s_q$ rays of type~4.

We now count.
The rays of type~1 and the partners of the rays of type~2a are unproductive, and they are distinct, since the former are charged and the latter are not.
So there are at most $\lambda$ rays of types~1 and~2a.
Map every ray of type~2b or~3 to a Steiner point passed by its component, chosen as above.
This point determines the component, since at most one component passes it.
For type~2b the component lies on a line that is not a vertex-pair line, and for type~3 it does, so the two types use distinct points.
A component of type~2b contains a single vertex, so it has only one productive end ray.
Two rays of type~3 with the same first corner $x$ would lie at the two ends of the component, which then would contain no corner other than $x$ in its relative interior, hence no vertex and no productive ray.
So the map is injective, and there are at most $|S|$ rays of types~2b and~3.
Finally, there are at most $\sum_{q\in S\setminus M}s_q$ rays of type~4.
\end{proof}

\begin{lemma}
\label{lem:rhostar}
$\rho_\star\le\rho+\lambda+|S|$.
\end{lemma}
\begin{proof}
By \cref{lem:links,lem:keep,lem:charged}, $\Psi(M)$ is a partition in $\K_\star$ with $|\Psi(M)|\le\rho+\chi-\sum_{q\in S\setminus M}s_q\le\rho+\lambda+|S|$.
\end{proof}

\subsection{The dynamic program}
\label{sec:dp}

We now show how to compute, in $O(n^{12})$ time, a partition with at most $\rho_\star$ triangles.
The algorithm extends the dynamic program for Steiner-free partitions~\cite{AAP86}: a state is a pair of \emph{terminals} instead of a pair of vertices, where a terminal is a vertex of $P$ or a point of $\C$ attached to a vertex, and the segments from vertices to their attached points play the role of the links.
That every derivation yields a partition is a winding-number identity.
That every partition in $\K_\star$ is found follows from the duality between spanning trees of a plane graph and of its dual: the links and all but one of the pieces of $\partial P$ between consecutive corners form a spanning tree of the graph of the sides, so the triangles form a tree, and the derivation is read off this tree.

\begin{figure}[t]
\centering
\captionsetup[subfigure]{skip=2pt}
\newcommand{\dpchain}{
  \path (0,0) coordinate (wi) (0.45,-0.85) coordinate (c1) (1.2,-1.15) coordinate (c2) (2,-1.05) coordinate (c3) (2.6,-0.7) coordinate (c4) (3,0) coordinate (wj) (0.6,1) coordinate (a) (2.4,1) coordinate (b); }
\newcommand{\dpchaindraw}{
  \draw[thick] (wi)--(c1)--(c2)--(c3)--(c4)--(wj); \foreach \q in {wi,c1,c2,c3,c4,wj} \fill (\q) circle (1.6pt); \node[left] at (wi) {$v_{k_1}$}; \node[right] at (wj) {$v_{k_2}$}; } \tikzset{steiner/.style={fill=myAlert}, dseg/.style={->,thick,shorten >=2pt}}
\begin{subfigure}[b]{0.4\textwidth}
\centering
\begin{tikzpicture}[scale=1]
  \dpchain
  \fill[myFill] (wi)--(c1)--(c2)--(c3)--(c4)--(wj)--(b)--(a)--cycle;
  \draw (wj)--(b) (a)--(wi); \draw[dseg] (b)--(a);
  \dpchaindraw
  \fill[steiner] (a) circle (2pt) node[above] {$p_1$};
  \fill[steiner] (b) circle (2pt) node[above] {$p_2$};
\end{tikzpicture}
\caption{}
\label{fig:states}
\end{subfigure}
\hfil
\begin{subfigure}[b]{0.4\textwidth}
\centering
\begin{tikzpicture}[scale=1]
  \dpchain
  \coordinate (r) at (1.5,0.15);
  \fill[myFill] (wi)--(c1)--(c2)--(c3)--(c4)--(wj)--(b)--(a)--cycle;
  \fill[myDark] (b)--(a)--(r)--cycle;
  \draw (wj)--(b)--(a)--(wi) (a)--(r)--(b);
  \draw[very thick,myAlert] (r)--(c3);
  \dpchaindraw
  \node[below] at (c3) {$v_{k_3}$};
  \fill[steiner] (a) circle (2pt) node[above] {$p_1$};
  \fill[steiner] (b) circle (2pt) node[above] {$p_2$};
  \fill[steiner] (r) circle (2pt) node[below left] {$p_3$};
\end{tikzpicture}
\caption{}
\label{fig:transition}
\end{subfigure}
\caption{(a)~A state $X(t_1,t_2)$ with $t_1=(k_1,p_1)$ and $t_2=(k_2,p_2)$; the arrow is the segment $p_2p_1$.
  (b)~A transition with $t_3=(k_3,p_3)$: the triangle $p_2p_1p_3$ is dark, and the two resulting states traverse the segment $v_{k_3}p_3$ (red) in opposite directions.}
\end{figure}
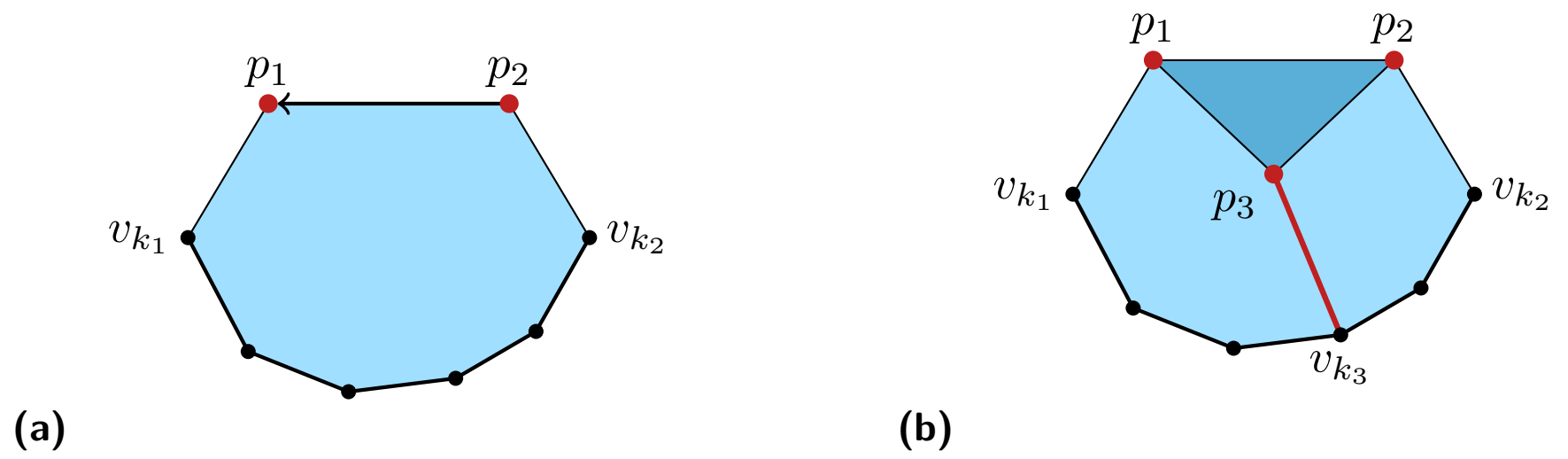

\subparagraph{Terminals and states.} Let $v_0,\dots,v_{n-1}$ be the vertices of $P$ in counterclockwise order, and $v_n=v_0$.
A \emph{terminal} is a pair $t=(k,p)$ with $0\le k\le n$ such that $p=v_k$, or $k<n$ and $p\in\C$ lies in the relative interior of the edge $v_kv_{k+1}$, or $k<n$ and $p\in\C$ lies in the interior of $P$ on a vertex-pair line through $v_k$.
We say that $p$ is \emph{attached} to $v_k$.
An edge of $P$ contains $O(n^2)$ points of $\C$, and so does each of the $O(n)$ vertex-pair lines through $v_k$, so there are $O(n^4)$ terminals.
We treat closed polygonal curves as formal sums of directed segments, modulo $xy+yx=0$ and $xy=xz+zy$ for $z\in xy$ (that is, as $1$-chains), and write $\omega_\gamma$ for the winding number of $\gamma$; it is defined outside $\gamma$ and additive under sums.
For $i\le j$, let $\gamma_{ij}$ be the chain $v_iv_{i+1}+\dots+v_{j-1}v_j$.
A \emph{state} $X(t_1,t_2)$ is an ordered pair of distinct terminals $t_1=(k_1,p_1)$ and $t_2=(k_2,p_2)$ with $k_1\le k_2$.
Its curve is (\cref{fig:states})
\[
\gamma_X=\gamma_{k_1k_2}+v_{k_2}p_2+p_2p_1+p_1v_{k_1}.
\]
The curve may be degenerate or self-intersecting; we use only the curve, never a region bounded by it.
The state is a \emph{leaf} if (L1) $k_1=k_2$ and $v_{k_1}$, $p_1$, $p_2$ are collinear, or (L2) $k_2=k_1+1$, $p_2=v_{k_2}$, and $p_1$ lies on the edge $v_{k_1}v_{k_2}$.
In both cases $\gamma_X=0$.

\subparagraph{Transitions.} A \emph{transition} of $X=X(t_1,t_2)$ is given by a terminal $t_3=(k_3,p_3)$ different from $t_1$ and $t_2$ with $k_1\le k_3\le k_2$, such that the triple $p_2p_1p_3$ is counterclockwise or collinear; a triple with a repeated point counts as collinear.
Its cost is $1$ or $0$, respectively.
It leads to the states $Y_1=X(t_1,t_3)$ and $Y_2=X(t_3,t_2)$ (\cref{fig:transition}), and
\begin{equation}
\label{eq:dpsplit}
\gamma_X=\partial(p_2p_1p_3)+\gamma_{Y_1}+\gamma_{Y_2},
\end{equation}
since $\gamma_{k_1k_3}+\gamma_{k_3k_2}=\gamma_{k_1k_2}$, and the segments $v_{k_3}p_3$, $p_1p_3$, and $p_3p_2$ each appear once in each direction and cancel.
A \emph{derivation} of $X$ is a finite tree of transitions rooted at $X$ whose leaves are leaf states.
Its cost is the sum of the costs of its transitions, and $D(X)\in\mathbb{N}\cup\{\infty\}$ is the minimum cost of a derivation of $X$.
The root state is $X_0=X((0,v_0),(n,v_n))$, and $\gamma_{X_0}=\partial P$.

\begin{lemma}[Soundness]
\label{lem:dpsound}
If $D(X_0)<\infty$, then a derivation of cost $D(X_0)$ yields a triangle partition of $P$ with $D(X_0)$ triangles, all of whose corners lie in $V\cup\C$.
\end{lemma}
\begin{proof}
The winding number of a counterclockwise triangle is $1$ on its interior and $0$ outside, and a collinear triple bounds the curve $0$.
Applying \cref{eq:dpsplit} along the derivation, its $D(X_0)$ counterclockwise triangles $\Delta_\ell$ satisfy $\sum_\ell\mathbf 1_{\Delta_\ell}=\omega_{\partial P}=\mathbf 1_P$ almost everywhere.
Hence their interiors are disjoint, their union is $P$, and their corners are points of terminals.
\end{proof}

\subparagraph{Evaluation.} The values $D(X)$ satisfy $D(X)=0$ for leaves and $D(X)=\min\{\operatorname{cost}+D(Y_1)+D(Y_2)\}$ for other states, where the minimum is over the transitions of $X$.
A transition with $k_3\in\{k_1,k_2\}$ leads to a state on the same pair of indices, and there are cycles of zero-cost transitions, so we do not process the states in a fixed order.
Instead we use Knuth's generalization of Dijkstra's algorithm~\cite{Knuth77}.
It applies since the value $\operatorname{cost}+D(Y_1)+D(Y_2)$ of a transition is nondecreasing in $D(Y_1)$ and $D(Y_2)$ and at least each of them.
The algorithm finalizes the states in nondecreasing order of value.
When it finalizes a state $X(t_a,t_b)$, it updates every state $X(t_a,t_c)$ for which $X(t_b,t_c)$ is final, and every state $X(t_c,t_b)$ for which $X(t_c,t_a)$ is final, using the transitions with $t_3=t_b$ and $t_3=t_a$, respectively.
When it finalizes a state, it stores a transition attaining its value, whose children are already final; these choices form a derivation of cost $D(X_0)$.
The values are integers, and $D(X_0)\le n-2$ by \cref{lem:dpcomplete} below, since triangulations belong to $\K_\star$; so larger values can be ignored, and a bucket queue takes amortized $O(1)$ time per operation.
To output the triangles, we expand the derivation from $X_0$ without entering states of value $0$, which contain no transition of cost $1$.
At each depth the values of the expanded states add up to at most $D(X_0)$, and the depth is at most the number of states, so this takes $O(n^9)$ time.

\subparagraph{Every partition in the class is found.} Fix a partition $\T'\in\K_\star$, not necessarily optimal, and choose a link $ru_r$ for every Steiner point $r$ of $\T'$ in the interior of $P$.
The links and all but one of the pieces of $\partial P$ between consecutive corners form a spanning tree of the plane graph of the sides of $\T'$, so the other sides form a spanning tree of its dual graph, whose nodes are the triangles and the exterior of $P$; rooted at the exterior, this tree is read off as a derivation with one transition of cost $1$ per triangle.

Let $Q$ be the set of corners of $\T'$, which contains $V$ by \cref{lem:corner}, and let $E$ be the union of the sides of $\T'$.
Every point of $E\setminus Q$ lies in the relative interior of every side that contains it, and these sides lie on one line, since sides do not cross.
Hence the closures of the connected components of $E\setminus Q$ are segments, and they are the edges of a plane graph $G$ with vertex set $Q$.
The connected components of the complement of $G$ are the interiors of the triangles of $\T'$ and the complement of $P$, which we call the \emph{exterior}.
The graph $G$ is connected, since two triangles that intersect have intersecting boundaries, and $P$, the union of the triangles, is connected.
An edge of $G$ on $\partial P$ has the exterior on one side and a triangle on the other, which lies to its left when the edge is directed counterclockwise along $\partial P$.
Every other edge has a triangle on each side.

Let $v_0=b_0,b_1,\dots,\bar b$ be the corners on $\partial P$ in counterclockwise order.
The \emph{boundary edges} are the edges $b_ib_{i+1}$ between consecutive corners in this order; the remaining edge $\bar bv_0$ of $G$ on $\partial P$ is the \emph{root edge}, and it is not a boundary edge.
Every chosen link is an edge of $G$, since it lies in $E$, its endpoints are corners, and no corner lies in its relative interior.
Every corner not on $\partial P$ is a Steiner point $r$ in the interior of $P$, and it is an endpoint of exactly one chosen link, whose other endpoint is a vertex of $P$.
So the boundary edges and the chosen links form a spanning tree $B$ of $G$: the path $b_0\dots\bar b$ with pendant edges.
The edges of $G$ not in $B$ are the root edge and the \emph{free edges}.

For a corner $q$, let $\beta(q)$ be $q$ if $q\in\partial P$, and $u_q$ otherwise, and let $\iota(q)$ be the index with $\beta(q)=b_{\iota(q)}$.
Let $k(q)$ be the index $k<n$ such that $\beta(q)$ lies on the edge $v_kv_{k+1}$ and differs from $v_{k+1}$, and let $t_q=(k(q),q)$.
By (i) and the definition of a link, $t_q$ is a terminal, and distinct corners give distinct terminals.
Moreover, $k(q)$ is nondecreasing in $\iota(q)$.
\begin{itemize}
\item (O1) \emph{If $xy$ is a boundary edge directed counterclockwise, or a chosen link directed either way, then $k(x)\le k(y)$ and $X(t_x,t_y)$ is a leaf.}
  For a link between $r$ and $v_k$, $k(x)=k(y)=k$, and (L1) holds since one of $x,y$ is $v_k$.
  For a boundary edge in which $y$ is not a vertex, $x$ and $y$ lie on the same edge $v_kv_{k+1}$, $k(x)=k(y)=k$, and $v_k,x,y$ are collinear, so (L1) holds.
  For a boundary edge with $y=v_{k+1}$, $k(x)=k$, $k(y)=k+1$, and $x$ lies on the edge $v_kv_{k+1}$, so (L2) holds.
\end{itemize}

Let $T^*$ be the graph whose nodes are the triangles of $\T'$ and the exterior, with one edge for each edge of $G$ not in $B$, joining the two nodes on its sides.
Then $T^*$ is a tree, by the classical duality between spanning trees and their complements: it is connected, since the complement of the tree $B$ in the plane is connected, so any two of these regions are joined by a polygonal path that avoids $B$ and $Q$ and crosses edges of $G$ finitely often; and it has $|Q|-1$ fewer edges than $G$, so, by Euler's formula, one edge fewer than nodes.
The exterior is incident only to the root edge among the edges not in $B$, so it has degree~$1$ in $T^*$, and we root $T^*$ at it.
For an edge $d$ of $T^*$, that is, the root edge or a free edge, let $\Delta_d$ be the triangle on the side of $d$ away from the root, and $\T_d$ the set of triangles in the subtree of $\Delta_d$.
Direct $d$ as $yx$ with $\Delta_d$ to its left.

\begin{lemma}
\label{lem:dpcomplete}
$D(X_0)\le|\T'|$ for every partition $\T'\in\K_\star$.
\end{lemma}
\begin{proof}
We use the notation above and prove two claims.

\emph{Claim 1: $k(x)\le k(y)$ for every edge $d=yx$ of $T^*$.}
For the root edge, $x=v_0$ and $k(x)=0$.
Let $d$ be a free edge, and let $\gamma_d=\sum_{\Delta\in\T_d}\partial\Delta$, a $1$-chain on the edges of $G$.
The coefficient of a directed edge in $\gamma_d$ is $[\text{the triangle to its left is in $\T_d$}]-[\text{the triangle to its right is in $\T_d$}]$.
For a free edge other than $d$, the two triangles on its sides are adjacent in $T^*-d$, so both or neither are in $\T_d$, and the coefficient is $0$.
For the root edge, the triangle to its left is the child of the root, which lies in $\T_d$ only if $d$ is the root edge, so the coefficient is $0$.
The coefficient of $d=yx$ is $1$, and the coefficient of a boundary edge directed counterclockwise is $0$ or $1$, since the exterior lies to its right.
So $\gamma_d=yx+\gamma'$, where $\gamma'$ is a chain on $B$ whose coefficients on the boundary edges directed counterclockwise are $0$ or $1$.
Since $\gamma_d$ is a sum of closed curves, the boundary of $\gamma'$ is $y-x$, and since $B$ is a tree, $\gamma'$ is the path from $x$ to $y$ in $B$.
This path runs along the boundary path from $b_{\iota(x)}$ to $b_{\iota(y)}$.
If $\iota(x)>\iota(y)$, it would traverse boundary edges clockwise, with coefficient $-1$ on their counterclockwise direction.
Hence $\iota(x)\le\iota(y)$ and $k(x)\le k(y)$.

\emph{Claim 2: $X_d=X(t_x,t_y)$ is a state and $D(X_d)\le|\T_d|$ for every edge $d=yx$ of $T^*$.}
It is a state by Claim~1 and since distinct corners give distinct terminals.
We use induction on $|\T_d|$.
Let $x=q_0,q_1,\dots,q_I=y$ be the corners on $\partial\Delta_d$ in counterclockwise order, so that $d$ is the edge from $q_I$ to $q_0$ and every $q_iq_{i+1}$ is an edge of $G$ with $\Delta_d$ to its left.
If $q_iq_{i+1}$ is in $B$, it is a link or a boundary edge directed counterclockwise, so by (O1), $k(q_i)\le k(q_{i+1})$ and $X(t_{q_i},t_{q_{i+1}})$ is a leaf.
Otherwise $q_iq_{i+1}$ is a free edge $d'$, not the root edge.
Indeed, the only triangle on the root edge is the child of the root, and this triangle is $\Delta_d$ only if $d$ is the root edge.
The triangle on the other side of $d'$ is a child of $\Delta_d$ in $T^*$, so it is $\Delta_{d'}$, it lies to the left of $q_{i+1}q_i$, and $X_{d'}=X(t_{q_i},t_{q_{i+1}})$.
By Claim~1, $k(q_i)\le k(q_{i+1})$, and by induction $D(X_{d'})\le|\T_{d'}|$.

Hence $k(q_0)\le k(q_1)\le\dots\le k(q_I)$, and $X(t_{q_i},t_{q_j})$ is a state for all $i<j$.
The points $q_0,\dots,q_I$ lie on the boundary of the convex set $\Delta_d$ in counterclockwise order, so for $i<j<l$ the triple $q_lq_iq_j$ is counterclockwise or collinear, and $t_3=t_{q_j}$ is a transition from $X(t_{q_i},t_{q_l})$ to $X(t_{q_i},t_{q_j})$ and $X(t_{q_j},t_{q_l})$.
Its cost is $1$ if $q_i,q_j,q_l$ are the corners of $\Delta_d$, and $0$ if they lie on one side of $\Delta_d$.
Triangulate the convex polygon with vertices $0,1,\dots,I$ by diagonals that include the pairs of indices of the corners of $\Delta_d$, if they are not polygon edges.
One triangle consists of the three corners, and every other triangle has its vertices on one side of $\Delta_d$.
Reading this triangulation from the polygon edge $\{0,I\}$ inwards gives a tree of transitions of total cost $1$ from $X_d$ to the states $X(t_{q_i},t_{q_{i+1}})$ for $i<I$.
Therefore $D(X_d)\le1+\sum_{d'}|\T_{d'}|=|\T_d|$, where $d'\ne d$ ranges over the free edges on $\partial\Delta_d$.

For the root edge $d=\bar bv_0$, we have $y=\bar b$ and $k(\bar b)=n-1$, and Claim~2 gives $D(X_d)\le|\T'|$.
The triple $v_nv_0\bar b$ has a repeated point, so $t_3=t_{\bar b}$ is a transition of cost $0$ from $X_0$ to $X_d$ and $X(t_{\bar b},(n,v_n))$, and the latter is a leaf by (L2).
Hence $D(X_0)\le|\T'|$.
\end{proof}

\begin{theorem}
\label{thm:dp}
A triangle partition of $P$ whose Steiner points each lie on two distinct lines through pairs of vertices of $P$ and that has at most $\rho_\star(P)$ triangles can be computed in $O(n^{12})$ time.
\end{theorem}
\begin{proof}
By \cref{lem:dpsound,lem:dpcomplete}, applied to an optimal partition in $\K_\star$, the derivation computed for $X_0$ yields a partition with Steiner points in $\C$ and $\rho(P)\le D(X_0)\le\rho_\star(P)$ triangles.
There are $O(n^4)$ terminals and $O(n^8)$ states.
Knuth's algorithm examines every triple of terminals $O(1)$ times, with one orientation test, so it takes $O(n^{12})$ time and $O(n^8)$ space.
The coordinates of the points of $\C$ are rational functions of constant degree of the input coordinates, so each test takes $O(1)$ exact arithmetic operations in the real-RAM model.
Computing $\C$ without duplicates, computing the terminals, and testing which points lie in $P$ take $O(n^5)$ time.
\end{proof}

\subsection{The approximation factor}
\label{sec:factor}

Combining the two constructions shows that $\rho_\star$ beats the factor $2$.
Intuitively, $\rho_0$ is close to $2\rho$ only if the slack $|S|+\lambda/2$ is small, and then $\rho_\star$ is close to $\rho$.

\begin{theorem}
\label{thm:53}
For every simple polygon $P$,
\[
\rho_\star(P)\le\frac{5\rho(P)-2}{3},
\]
where $\rho_\star(P)$ is the minimum size of a partition in $\K_\star$.
Hence the dynamic program of \cref{thm:dp} is a $\frac53$-approximation for $\rho$ that runs in $O(n^{12})$ time.
\end{theorem}
\begin{proof}
By $\rho_\star\le\rho_0$ and \cref{thm:main,lem:rhostar},
$3\rho_\star\le2\rho_0+\rho_\star\le2(2\rho-1-|S|-\lambda/2)+(\rho+\lambda+|S|)=5\rho-2-|S|\le5\rho-2$.
\end{proof}

\bibliography{refs}

\section*{AI Statement}
This paper was prepared with substantial help from the AI assistant Claude (Anthropic).
It took part in developing the algorithms and proofs, writing the code used to test them, checking the proofs and the references, drawing the figures, and writing the text.
The author checked all results and is responsible for the content.

\appendix
\crefalias{section}{appendix}
\crefalias{subsection}{appendix}

\section{Partitions versus covers}
\label{app:covers}

This appendix writes out the lower bound $\rho(P)/\kappa(P)=\Omega(\alpha(n))$ stated in the introduction.
It is the construction sketched in~\cite{Alk25}: a simple polygon that is the union of $N$ triangles can have $\Omega(N\,\alpha(N))$ edges.
A set of segments with a long lower envelope~\cite{WS88} is turned into a polygon that is covered by one triangle per segment.
We write out the details, and count reflex vertices instead of edges.

\begin{lemma}[{after~\cite{Alk25}}]
\label{lem:separation}
For infinitely many $N$ there are simple polygons with $\kappa(P)\le N+1$ and $\rho(P)=\Omega(N\,\alpha(N))$.
\end{lemma}
\begin{proof}
For a segment $\sigma$ and $x\in\mathbb{R}$, let $\sigma(x)$ be the smallest $y$ with $(x,y)\in\sigma$, and $\sigma(x)=+\infty$ if there is none.
Let $\Sigma$ be a finite set of segments.
We say that the lower envelope of $\Sigma$ has \emph{length} at least $\nu$ if there are $x_1<\dots<x_\nu$ and $\tau_1,\dots,\tau_\nu\in \Sigma$ with $\tau_i\ne\tau_{i+1}$, such that $x_i$ lies in the interior of the $x$-projection of $\tau_i$ and $\tau_i(x_i)<\sigma(x_i)$ for every $\sigma\in \Sigma\setminus\{\tau_i\}$.
There is a constant $\varepsilon>0$ such that, for infinitely many $N$, some set of $N$ segments has a lower envelope of length at least $\varepsilon N\,\alpha(N)$~\cite{WS88} (see also~\cite[Chapter~4]{SA95}).

Fix such a set $\Sigma$ with $|\Sigma|=N$.
The defining conditions survive small perturbations of the endpoints of the segments and of $x_1,\dots,x_\nu$: the point $x_i$ stays in the interior of the $x$-projection of $\tau_i$, so $\tau_i(x_i)$ changes continuously, and $\sigma(x_i)$ is lower semicontinuous in $x_i$ and in the endpoints of $\sigma$, since segments are closed, so the strict inequalities persist.
We may therefore assume that no segment of $\Sigma$ is vertical, no two are parallel, no three have a common point, and the endpoints of the segments, their pairwise intersection points, and $x_1,\dots,x_\nu$ all have pairwise distinct $x$-coordinates.
A map $(x,y)\mapsto(x,y')$, where $y'$ is an increasing affine function of $y$, preserves these properties, so we may also assume $\Sigma\subset(-W,W)\times(0,1)$ for some $W>0$.
Let $\bar\sigma=[-W,W]\times\{1\}$ and $\bar\Sigma=\Sigma\cup\{\bar\sigma\}$.

\subparagraph{The polygon.} Let $A=(0,1)$ and $f(x,y)=\bigl(\frac{x}{1+y},\frac{y}{1+y}\bigr)$.
This is a projective map, and a bijection from the half-plane $y>-1$ onto the half-plane $y<1$, with inverse $f^{-1}(x,y)=\bigl(\frac{x}{1-y},\frac{y}{1-y}\bigr)$.
It maps segments to segments, and it maps the upward vertical ray from a point $q$ onto the segment $f(q)A$ minus the point $A$.
For $\sigma\in \bar\Sigma$ let $T_\sigma=\operatorname{conv}(f(\sigma)\cup\{A\})$, a nondegenerate triangle because $\sigma$ is not vertical, and let
\[
P=\bigcup_{\sigma\in \bar\Sigma}T_\sigma .
\]
Then $\kappa(P)\le N+1$.
The triangle $T_\sigma$ is the image of the vertical half-strip $H_\sigma=\{(x,y):\sigma(x)\le y\}$, that is, $T_\sigma\setminus\{A\}=f(H_\sigma)$.
Hence $P\setminus\{A\}=f(U)$, where
\[
U=\bigcup_{\sigma\in \bar\Sigma}H_\sigma=\{(x,y):-W\le x\le W,\ y\ge \env(x)\},\qquad \env(x)=\min_{\sigma\in \bar\Sigma}\sigma(x)\le 1 .
\]
Let $-W=\xi_0<\xi_1<\dots<\xi_J=W$ be $\pm W$ together with the $x$-coordinates of the endpoints and pairwise intersection points of the segments of $\Sigma$.
On each interval $(\xi_{j-1},\xi_j)$ the set of segments of $\bar\Sigma$ whose $x$-projection contains it is fixed, and no two of them cross.
Since no two are parallel and $\bar\sigma$ lies strictly above every segment of $\Sigma$, a single segment $\sigma_j\in \bar\Sigma$ attains $\env$ on the whole interval.
Thus $\partial U$ consists of the two rays $\{\pm W\}\times[1,\infty)$ and an $x$-monotone polygonal path $\Gamma$ from $(-W,1)$ to $(W,1)$, formed by the graphs of $\sigma_j$ over $[\xi_{j-1},\xi_j]$, where consecutive graphs are joined by a vertical segment (possibly a point) at $\xi_j$.
Since $f$ is a homeomorphism that maps vertical lines into lines through $A$, $P$ is the closed region bounded by the polygonal curve $f(\Gamma)\cup f(W,1)A\cup Af(-W,1)$.
This curve is simple: $f(\Gamma)$ is simple because $f$ is injective, and the two segments at $A$ lie on the images of the lines $x=\pm W$, which meet $\Gamma$ only at $(\pm W,1)$.
Hence $P$ is a simple polygon, and $A$ lies in its kernel (\cref{fig:curtains}).

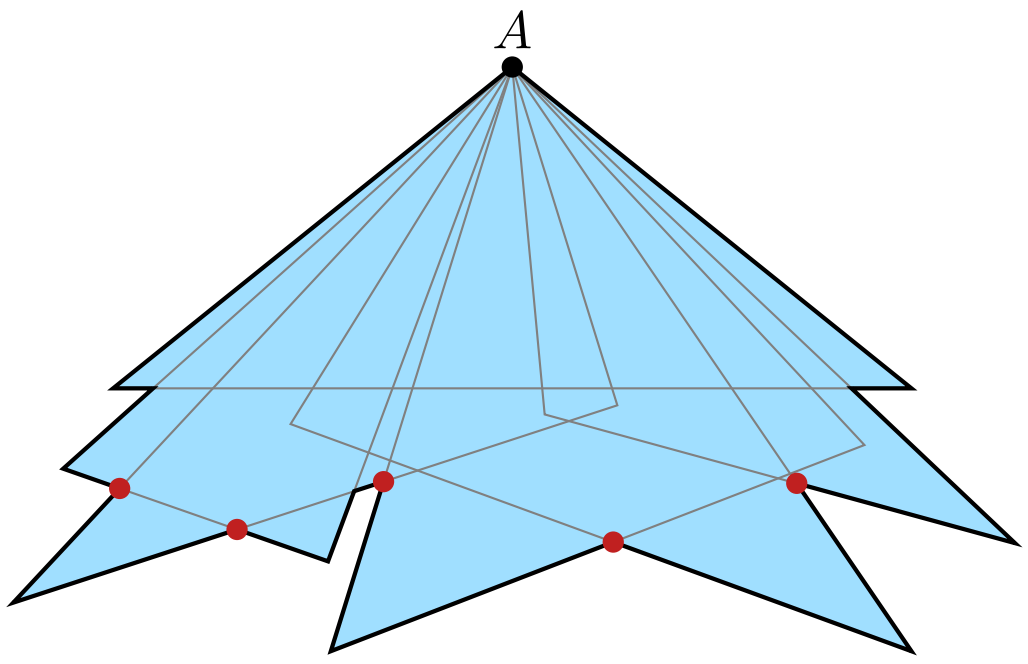
\begin{figure}[t]
\centering
\begin{tikzpicture}[x=0.6cm,y=0.85cm]
  \fill[myFill] (-1.419,1.774) -- (-1.743,1.7) -- (-2.031,1.154) -- (-3.034,1.402) -- (-5.5,0.833) -- (-4.328,1.721) -- (-4.95,1.875) -- (-3.96,2.5) -- (-4.4,2.5) -- (0,5) -- (4.4,2.5) -- (3.74,2.5) -- (5.541,1.296) -- (3.136,1.761) -- (4.4,0.455) -- (1.113,1.304) -- (-2,0.455) -- cycle;
  \draw[gray] (-5.5,0.833) -- (1.158,2.368) -- (0,5) -- cycle;
  \draw[gray] (-2.444,2.222) -- (4.4,0.455) -- (0,5) -- cycle;
  \draw[gray] (-2,0.455) -- (3.882,2.059) -- (0,5) -- cycle;
  \draw[gray] (0.357,2.297) -- (5.541,1.296) -- (0,5) -- cycle;
  \draw[gray] (-4.95,1.875) -- (-2.031,1.154) -- (0,5) -- cycle;
  \draw[gray] (-4.4,2.5) -- (4.4,2.5) -- (0,5) -- cycle;
  \draw[thick] (-1.419,1.774) -- (-1.743,1.7) -- (-2.031,1.154) -- (-3.034,1.402) -- (-5.5,0.833) -- (-4.328,1.721) -- (-4.95,1.875) -- (-3.96,2.5) -- (-4.4,2.5) -- (0,5) -- (4.4,2.5) -- (3.74,2.5) -- (5.541,1.296) -- (3.136,1.761) -- (4.4,0.455) -- (1.113,1.304) -- (-2,0.455) -- cycle;
  \foreach \q in {(-1.419,1.774),(-3.034,1.402),(-4.328,1.721),(1.113,1.304),(3.136,1.761)} \fill[myAlert] \q circle (2pt);
  \fill (0,5) circle (2pt) node[above] {$A$};
\end{tikzpicture}
\caption{The polygon $P$ built from five segments.
  The triangles $T_\sigma$ are drawn in gray; the top one is $T_{\bar\sigma}$.
  The red vertices come from crossings on the lower envelope and are reflex.}
\label{fig:curtains}
\end{figure}

\subparagraph{Reflex vertices.} Let $0<j<J$ be such that $\sigma_j\ne \sigma_{j+1}$ and $\xi_j$ is the $x$-coordinate of an intersection point of two segments of $\Sigma$.
Then $\xi_j$ is not the $x$-coordinate of an endpoint, so $\env$ is continuous at $\xi_j$ and both $\sigma_j$ and $\sigma_{j+1}$ contain $p=(\xi_j,\env(\xi_j))$.
No other segment contains $p$, so near $\xi_j$ every other segment lies strictly above $p$, and near $p$ the set $U$ is the union of the closed half-planes above the lines of $\sigma_j$ and $\sigma_{j+1}$.
These lines are distinct and cross at $p$, so $p$ is a vertex of $U$ with interior angle larger than $\pi$.
The map $f$ is a local diffeomorphism at $p$ that maps lines to lines, so $f(p)$ is a reflex vertex of $P$.

Now let $1\le i<\nu$.
The segment $\tau_i$ is the unique minimizer at $x_i$ (also against $\bar\sigma$, since $\tau_i(x_i)<1$), so $\tau_i=\sigma_j$ for the interval $(\xi_{j-1},\xi_j)$ containing $x_i$, and similarly for $\tau_{i+1}$.
As $\tau_i\ne\tau_{i+1}$, some $\xi_j\in(x_i,x_{i+1})$ has $\sigma_j\ne \sigma_{j+1}$.
If no endpoint of a segment of $\Sigma$ has its $x$-coordinate in $(x_i,x_{i+1})$, then $\xi_j$ is the $x$-coordinate of an intersection point, and $P$ has a reflex vertex $f(p)$ with $p$ in the slab $(x_i,x_{i+1})\times\mathbb{R}$.
These slabs are disjoint, and at most $2N$ of them contain the $x$-coordinate of an endpoint.
Therefore $P$ has at least $\nu-1-2N\ge \varepsilon N\,\alpha(N)-2N-1$ reflex vertices.
By \cref{lem:corner}, every partition of $P$ into triangles has at least a third of this number of triangles, and $\rho(P)=\Omega(N\,\alpha(N))$ because $\alpha(N)\to\infty$.
By the upper bound in the introduction~\cite{PSS88,GSS89}, $P$ has $n=O(N\,\alpha(N))$ vertices, so $\rho(P)/\kappa(P)=\Theta(\alpha(n))$ for these polygons.
\end{proof}

Consequently, a triangulation approximates $\kappa$ within a factor $O(\alpha(n))$, but no algorithm that outputs a partition, with or without Steiner points, approximates $\kappa$ within a constant factor.
Constant-factor algorithms for triangle covers~\cite{KS27} must therefore output overlapping triangles.

\end{document}